\documentclass[final,5p,times,twocolumn]{elsarticle}

\usepackage{amssymb}
\usepackage{amsthm,amssymb, amsfonts, amsmath}
\usepackage{units}
\usepackage{graphicx}
\usepackage{listings}
\usepackage{multirow}
\usepackage{booktabs}
\usepackage{float}
\floatstyle{plaintop}
\restylefloat{table}
\usepackage{booktabs}
\usepackage{enumerate}
\graphicspath{{figures/}}
\usepackage{caption}
\usepackage{subfig}
\usepackage{xcolor}

\journal{Annals of Nuclear Energy}

\begin{document}

\begin{frontmatter}



\title{On the Feynman-alpha Method for Reflected Fissile Assemblies}


\author[1,2]{Michael Y. Hua}
\author[2]{Jesson D. Hutchinson}
\author[2]{George E. McKenzie}
\author[1]{Shaun D. Clarke}
\author[1]{Sara A. Pozzi}

\address[1]{Department of Nuclear Engineering and Radiological Sciences, University of Michigan, Ann Arbor, MI 48109}
\address[2]{NEN-2: Advanced Nuclear Technology Group, Los Alamos National Laboratory, Los Alamos, NM 87545}

\begin{abstract}	
The Feynman-alpha method is a neutron noise technique that is used to estimate the prompt neutron period of fissile assemblies. The method and quantity are of widespread interest including in applications such as nuclear criticality safety, safeguards and nonproliferation, and stockpile stewardship; the prompt neutron period may also be used to infer the $k_\text{eff}$ multiplication factor. The Feynman-alpha method is predicated on time-correlated neutron detections that deviate from a Poisson random variable due to multiplication. Traditionally, such measurements are diagnosed with one-region point kinetics, but two-region models are required when the fissile assembly is reflected. This paper presents a derivation of the two-region point kinetics Feynman-alpha equations based on a double integration of the Rossi-alpha equations, develops novel propagation of measurement uncertainty, and validates the theory. Validation is achieved with organic scintillator measurements of weapons-grade plutonium reflected by various amounts of copper to achieve $k_\text{eff}$ values of 0.83-0.94 and prompt periods of 5-75 ns.  The results demonstrate that Feynman-alpha measurements should use the two-region model instead of the one-region model. The simplified one-region model deviates from the validated two-region models by as much as 10\% in the estimate of the prompt neutron period, and the two-region model reduces to the one-region model for small amounts of reflector. The Feynman-alpha estimates of the prompt neutron period are compared to those of the Rossi-alpha approach.  The comparative results demonstrate that the Feynman-alpha method is more precise than the Rossi-alpha method and more accurate for $k_\text{eff}<0.92$, whereas the Rossi-alpha method is generally more accurate for higher multiplications.  The uncertainty propagation developed in this work should be used for all Feynman-alpha measurements and will therein improve fitting accuracy and appropriate precision estimates.
\end{abstract}


\end{frontmatter}

\section{Introduction}\label{sec:introduction}
The measurement of nuclear material is of broad interest with applications in nuclear criticality safety, safeguards and nonproliferation, and stockpile stewardship for examples~\cite{Doyle}. One such measurement is the Feynman-alpha method, which utilizes time correlations between neutron detections to estimate the prompt neutron period of fissile assemblies~\cite{Feynman44_1,Feynman44_2,Feynman56,uhrig}.  The prompt neutron period is itself interesting in the context of applications and it is also used to infer derivative values such as the $k_\text{eff}$ multiplication factor.  

The Feynman-alpha method was originally developed for bare cores of fissile material and shortcomings in adequacy have been identified when reflectors are introduced~\cite{Jesson2017}. Therefore, it is desirable to extend the traditional one-region point kinetic models to two-region point kinetics to account for the region introduced by reflector, which has been addressed in greater generality by existing literature~\cite{Chernikova2014,munoz_cobo}. This work simplifies the equations and follows a different derivation based on the double integration of the Rossi-alpha method~\cite{double_int,double_int2}.  The Rossi-alpha-integration approach is selected to utilize recent two-region point kinetic generalizations to the Rossi-alpha equations~\cite{mikwa_2exp}, which have been validated~\cite{mikwa_validation}.

Beyond the derivation, this work develops rigorous first-order propagation of measurement uncertainty similar to Refs.~\cite{mikwa_unc,mikwa_ANS_unc}. The propagation involves weighting a fit to the Feynman histogram data, which results in a more accurate fit that adequately incorporates measurement uncertainty with fit uncertainty to estimate a composite uncertainty in the fit parameters.  The prompt neutron period is a function of the fit parameters in the two-region model (whereas it is a fit parameter in the one-region model), thus the uncertainty in fit parameters is propagated to a final estimate of the prompt neutron period. Two methods of determining the uncertainty in the Feynman histogram data, analytic and sample methods, are presented.

The theory is validated with experimental data and simulations to determine reference values.  An organic scintillator array (OSCAR) is used to measure a 4.5-kg sphere of weapons-grade, alpha-phase plutonium reflected by various amounts of copper to attain different levels of reflection and multiplication.  Additionally, the one- and two-region estimates of the prompt neutron period are compared.

The structure of this paper is as follows.  Background on the Rossi-alpha method is provided in Sec.~\ref{sec:background}, and the derivation of the two-region Feynman-alpha theory based on the double integration of the Rossi-alpha equations is presented in Sec.~\ref{sec:theory}.  Associated uncertainty propagation is discussed in Sec.~\ref{sec:unc}, the experimental setups are shown in Sec.~\ref{sec:setup}, and the analysis of data is discussed in Sec.~\ref{sec:analysis}. Results are presented in Sec.~\ref{sec:results} and subsequent conclusions are made in Sec.~\ref{sec:conclusion}.

\section{Rossi-alpha Method}\label{sec:background}
The Rossi-alpha method is a neutron noise technique used to estimate the prompt neutron decay constant $\alpha$ (or its negative reciprocal, the prompt neutron period) of near- and delayed-critical assemblies of fissile material~\cite{Feynman44_1,Feynman44_2,Feynman56,uhrig}.  The method is based on the nonrandom distribution \textit{between} neutron detection times: the population of correlated prompt neutrons (those from the same fission chain seed) decays exponentially as fission chains extinguish~\cite{ornbro}.  Experimental results showed that the decay is modeled better by a two-exponential function than a one-exponential function when the assembly of fissile material is reflected~\cite{PHYSOR2006,Jesson2017}, the model was theoretically derived from two-region point kinetics~\cite{Avery,Cohn,mikwa_2exp}, and validated with a combination of experiments and simulations~\cite{mikwa_validation,RA_w_Organics}.  The two-exponential Rossi-alpha probability density function for reflected assemblies $p_R(\tau)$ is given by 

\begin{equation}\label{eq:exp2}
p_R(\tau)\ dt =A\ d\tau + B\left[\rho_1 e^{r_1 t} + \rho_2 e^{r_2 t}\right]\ d\tau, 
\end{equation}

where the $B$ term represents the correlated neutrons that decay to the $A={F_0}^2\varepsilon_g\varepsilon_c$ term that represents the uniform probability of uncorrelated, chance coincidences.  The $F_0$ variable is the average fission rate in the system.  There are two efficiency terms: $\varepsilon_c$ is the efficiency of the counting channel and $\varepsilon_g$ is the trigger channel efficiency.  The efficiencies may represent a single detector (autocorrelation, Rossi-alpha) or two separate detectors (cross correlation, Feynman-alpha); in this work, the terms are implicitly condensed to $\varepsilon^2 = \varepsilon_g\varepsilon_c$ ~\cite{double_int}.  The exponents $r_1$ and $r_2$ are related to physical parameters by

\begin{equation}\label{eq:r_simp}
r_j = (-1)^j \sqrt{\frac{1}{\tau_r}\left( f' + \alpha\right) + \frac{1}{4}\left(\frac{1}{\tau_r}-\alpha\right)^2} + \frac{1}{2}\left(\alpha-\frac{1}{\tau_r}\right), 
\end{equation}

where $\tau_r$ is the mean neutron lifetime in the reflector region and $f' = f_{cr}f_{rc}/\tau_c$ is a function of cross-region leakage terms $f_{cr}$ and $f_{rc}$ and the mean neutron lifetime in the fissile core $\tau_c$.  The coefficients to the exponential terms are given by

\begin{subequations}\label{eq:rho}
	\begin{equation}
	\rho_1 = \frac{(1-R)^2}{r_1} + \frac{2(1-R)(R)}{r_1+r_2} \label{eq:rho1}\\
	\end{equation}
	and
	\begin{equation} 
	\rho_2 = \frac{R^2}{r_2} + \frac{2(1-R)(R)}{r_1+r_2}, \label{eq:rho2}
	\end{equation}
\end{subequations}

where $R$ is a weighting parameter used to calculate 

\begin{equation}
\alpha = r_1(1-R) + r_2(R).\label{eq:alpha}
\end{equation}

Work validating the two-region Rossi-alpha model demonstrated improved agreement as the $k_\text{eff}$ multiplication factor tends to unity~\cite{mikwa_validation}.

\section{Feynman-alpha Method}\label{sec:theory}
The number of neutron counts $c$ in a given window $\tau$ deviates from a Poisson random variable due to multiplication/fluctuations in the prompt neutron population~\cite{Feynman44_1,Feynman44_2,Feynman56, fluctuations}.  The deviation is used as a signature to estimate the prompt neutron decay constant $\alpha$; the signature $Y$ is called ``excess variance" and is related to $c$ by
\begin{equation}
\label{eq:Yc}
Y = \frac{\text{variance}}{\text{mean}}-1 = \frac{\langle c^2 \rangle - \langle c \rangle^2}{\langle c\rangle}-1,
\end{equation}
where $\langle c^2\rangle$ is the second moment of the counting distribution for a given window $\tau$. In the one-region point kinetics model, $Y$ is related to $\alpha$ by 
\begin{equation}\label{eq:Ya_one}
Y = \gamma\left(1-\frac{1-e^{-\alpha\tau}}{\alpha\tau}\right),
\end{equation}
where $\gamma$ is a scaling constant and both $\gamma$ and $\alpha$ are determined by fitting $Y$ as a function of $\tau$.  The probability density function for the two-region Rossi-alpha distribution in Eqn.~\eqref{eq:exp2} can be integrated twice and algebraically manipulated~\cite{double_int,munoz_cobo,double_int2} to obtain the two-region point kinetics model.  Note that other methods of derivation exist in greater generality~\cite{Chernikova2014}.  The double integration is related to $c$ by
\begin{equation}\label{eq:c_rossi_int}
\frac{\langle c^2\rangle-\langle c\rangle}{2} = \int_{0}^{\tau} dt_2 \int_{0}^{t2} dt_1 p_{\text{R}}(t_2 - t_1),
\end{equation}
where $\tau = t_2 - t_1$.  Performing the integrations results in
\begin{equation}
\frac{\langle c^2\rangle-\langle c\rangle}{2} = \frac{{F_0}^2\epsilon^2\tau^2}{2} - B'F_0\epsilon^2\tau\left(\frac{\rho_1}{{r_1}}\left(1+\frac{1-e^{r_1\tau}}{r_1\tau}\right) +\frac{\rho_2}{{r_2}}\left(1+\frac{1-e^{r_2\tau}}{r_2\tau}\right) \right),
\end{equation}
where $B'$ is treated as an arbitrary constant.  Noting that $\langle c\rangle = F_0\epsilon\tau$, multiplying both sides of the equation by $2/\langle c\rangle$, and rearranging terms further results in
\begin{equation}
Y=\frac{\langle c^2\rangle-\langle c\rangle^2}{\langle c\rangle}-1 = - B'\epsilon\left(\frac{\rho_1}{{r_1}}\left(1+\frac{1-e^{r_1\tau}}{r_1\tau}\right) +\frac{\rho_2}{{r_2}}\left(1+\frac{1-e^{r_2\tau}}{r_2\tau}\right) \right).
\end{equation}
The equation can be condensed into
\begin{subequations}\label{eq:Y2}
	\begin{equation}
		Y = \gamma_1\left(1+\frac{1-e^{r_1\tau}}{r_1\tau}\right) + \gamma_2\left(1+\frac{1-e^{r_2\tau}}{r_2\tau}\right)
	\end{equation}
for practical fitting applications, where
	\begin{eqnarray}
		\gamma_1 &= -B''\frac{\rho_1}{r_1}\\
		\gamma_2 &= -B''\frac{\rho_2}{r_2}
	\end{eqnarray}
\end{subequations}
such that there are only four fit parameters: $\gamma_1,\gamma_2,r_1,$ and $r_2$. Note that $B''$ absorbs the efficiency variable to define a new arbitrary constant. The value of $R$ is calculated by taking the ratio of $\gamma_1$ and $\gamma_2$; the ratio eliminates $B''$ and results in an equation relating numeric values, $R$, $r_1$, and $r_2$.  Since $r_1$ and $r_2$ are known from the fit, $R$ may be calculated.  Then, the numeric values for $R$, $r_1$, and $r_2$ are used to calculate $\alpha$ by Eqn.~\eqref{eq:alpha}.  Hence, using Eqns.~\eqref{eq:rho},~\eqref{eq:alpha}, and~\eqref{eq:Y2}, $\alpha$ is expressed in terms of the fit parameters as:
\begin{equation}
\label{eq:alpha_fit}
\alpha = (r_1 + r_2) + r_1r_2\sqrt{\frac{\gamma_1+\gamma_2}{\gamma_1 {r_1}^2 + \gamma_2 {r_2}^2}}.
\end{equation}

\subsection{In Terms of Factorial Moments}
Equation~\eqref{eq:Y2} can be applied to the $Y_2$ parameter, which in turn can be expressed in terms of factorial counting moments by~\cite{hage_cifarelli_factMoment,UQ_jesson}
\begin{equation}\label{eq:Y_moment}
Y_2(\tau) = \frac{1}{2}\frac{m_1(\tau)}{\tau}Y(\tau) = \frac{1}{\tau}\left(m_2(\tau)-\frac{1}{2}{m_1}^2(\tau)\right).
\end{equation}
The factorial moments $m_2$ and $m_1$ are calculated as a function of $\tau$ from list mode data (a sorted list of neutron detection times) and the random gate generation technique is used in this work~\cite{hage_cifarelli_factMoment}. Mathematically,
\begin{equation}\label{eq:moment}
m_\mu(\tau) \approx \sum^{\infty}_{x=\mu} {x \choose \mu} {b_x}^{+}(\tau),
\end{equation}
where
\begin{equation}
{b_x}^{+}(\tau) = \frac{B_x(\tau)}{K}
\end{equation}
and $B_x(\tau)$ is, after looking in $K$ inspection windows, the number of windows containing exactly $x$ neutron detections.  The Feynman histogram is constructed by calculating $Y_2$ as a function of $\tau$. 

\section{Propagation of Measurement Uncertainty}\label{sec:unc}
Rigorous quantification of measurement uncertainty for the Feynman-alpha method that propagates histogram uncertainty (uncertainty in $Y_2$) through the fitting algorithm to $\alpha$ is needed.  Furthermore, whereas $\alpha$ is a fit parameter in the one-region model, it is a function of the fit parameters in the two-region model.  Therefore, uncertainty must be propagated from the fit parameters to the final estimate of $\alpha$.  Subsection~\ref{sec:unc_hist} presents two methodologies for estimating histogram uncertainty, subsec.~\ref{sec:unc_fit} describes the process for propagating histogram and fit uncertainty to the fit parameters, and subsec.~\ref{sec:unc_param} propagates uncertainty from the fit parameters to the final estimate of $\alpha$.  The collective process is adapted from Refs.~\cite{mikwa_unc, mikwa_ANS_unc}.

\subsection{Uncertainty in the Feynman Histogram}\label{sec:unc_hist}
The first method of determining histogram uncertainty, called the sample method, divides a total measurement into multiple smaller measurements, calculates a histogram for each submeasurement, and then calculates the error bars by taking bin-by-bin standard deviations. The second method -- an analytic approach -- calculates the bin-by-bin standard deviation using up to the fourth factorial moment (see Eqn.~\eqref{eq:moment}) and is given by~\cite{UQ_jesson,UQ_LLNL}
\begin{equation}
\sigma_{Y_2} = \frac{1}{\tau}\sqrt{\frac{1}{K-1}\left( 6m_4-6m_3m_1+6m_3-{m_2}^2+4m_2{m_1}^2-4m_2m_1+m_2-{m_1}^4+{m_1}^3\right) }.
\end{equation}
The analytic method is preferential to the sample method so long as the calculation of the fourth factorial moment is reliable.  Reliability depends on a variety of variables such as accidentals rate, detector efficiency, assembly multiplication, and measurement time; if the accidentals rate is comparable to the fourth factorial moment, the uncertainty in the latter may become unbounded.

\subsection{Uncertainty Propagation through Fitting Algorithm to Fit Parameters}\label{sec:unc_fit}
Weighting the nonlinear least-squares fit to experimental data reduces fit uncertainty, results in greater accuracy, and appropriately propagates experimental uncertainty through the fit algorithm.  Inverse-variance weighting -- weighting bin $i$ by $1/{\sigma_i}^2$ -- optimally reduces the uncertainty.  The uncertainties in the fit parameters are given by the covariance matrix,
\begin{equation}
\Sigma_\text{covar} = [J^T W J]^{-1} = 
	\begin{bmatrix}
		{\sigma_{\gamma_1}}^2		&\sigma_{\gamma_1\gamma_2}	&\sigma_{\gamma_1r_1}	&\sigma_{\gamma_1r_2} \\
		\sigma_{\gamma_2\gamma_1}	&{\sigma_{\gamma_2}}^2		&\sigma_{\gamma_2r_1}	&\sigma_{\gamma_2r_2} \\
		\sigma_{r_1\gamma_1}			&\sigma_{r_1\gamma_2}		&{\sigma_{r_1}}^2		&\sigma_{r_1r_2} \\
		\sigma_{r_2\gamma_1}			&\sigma_{r_2\gamma_2}		&\sigma_{r_2r_1}		&{\sigma_{r_2}}^2 \\
\end{bmatrix},
\end{equation} 
that has variances on the diagonal and covariances on the off-diagonal terms. The Jacobian matrix $J$ is an output of the fitting algorithm and an $[N\times P]$ matrix where $N$ is the number of histogram bins and $P$ is the number of fit parameters ($P=4$ when fitting with Eqn.~\eqref{eq:Y2}).  The weighting matrix $W$ is $[N\times N]$, zero on off-diagonal terms, $W_{i,i} = 1/{\sigma_i}^2$ on the diagonal, and given by
\begin{equation}
	W = \begin{bmatrix}
		\frac{1}{{\sigma_1}^2} &	&	& & \\
		&\frac{1}{{\sigma_2}^2} &	& &\\
		& &\ddots &\\
		& & & \frac{1}{{\sigma_N}^2}
	\end{bmatrix}.
\end{equation}

\subsection{Uncertainty Propagation from Fit Parameters to $\alpha$}\label{sec:unc_param}
The first order uncertainty in $\alpha$ is determined by propagating the uncertainty in the fit parameters through Eqn.~\eqref{eq:alpha_fit} via
\begin{align}\label{eq:var_alpha}
{\sigma_\alpha}^2 & = 
\left(\frac{\partial\alpha}{\partial r_1}\right)^2 {\sigma_{r_1}}^2 
+ \left(\frac{\partial\alpha}{\partial r_2}\right)^2 {\sigma_{r_2}}^2 
+ \left(\frac{\partial\alpha}{\partial \gamma_1}\right)^2 {\sigma_{\gamma_1}}^2 
+ \left(\frac{\partial\alpha}{\partial \gamma_2}\right)^2 {\sigma_{\gamma_2}}^2 \nonumber\\
&\quad 
+ 2\left(\frac{\partial\alpha}{\partial r_1}\right)
\left(\frac{\partial\alpha}{\partial r_2}\right) {\sigma_{r_1r_2}}  + 2\left(\frac{\partial\alpha}{\partial r_1}\right)
\left(\frac{\partial\alpha}{\partial \gamma_1}\right) {\sigma_{r_1\gamma_1}} 
+ 2\left(\frac{\partial\alpha}{\partial r_1}\right)
\left(\frac{\partial\alpha}{\partial \gamma_2}\right) {\sigma_{r_1\gamma_2}} \nonumber\\
&\quad
+ 2\left(\frac{\partial\alpha}{\partial r_2}\right)
\left(\frac{\partial\alpha}{\partial \gamma_1}\right) {\sigma_{r_2\gamma_1}} 
+ 2\left(\frac{\partial\alpha}{\partial r_2}\right)
\left(\frac{\partial\alpha}{\partial \gamma_2}\right) {\sigma_{r_2\gamma_2}} \nonumber\\
&\quad 
+ 2\left(\frac{\partial\alpha}{\partial \gamma_1}\right)
\left(\frac{\partial\alpha}{\partial \gamma_2}\right) {\sigma_{\gamma_1\gamma_2}} .
\end{align}
The partial derivatives are given by 
\begin{subequations}
\begin{align}
	\frac{\partial\alpha}{\partial\gamma_1} &= \frac{\gamma_2(\gamma_1+\gamma_2)(r_1r_2)^4({r_2}^2-{r_1}^2)}{2\delta^3}\\
	\frac{\partial\alpha}{\partial\gamma_2} &= \frac{\gamma_1(\gamma_1+\gamma_2)(r_1r_2)^4({r_1}^2-{r_2}^2)}{2\delta^3}\\
	\frac{\partial\alpha}{\partial r_1} &=\frac{(\gamma_1+\gamma_2)(r_1r_2)^2(\gamma_2{r_2}^4r_1(\gamma_1+\gamma_2)+\delta(\gamma_1{r_1}^2+\gamma_2{r_2}^2))}{\delta^3}\\
	\frac{\partial\alpha}{\partial r_2} &=\frac{(\gamma_1+\gamma_2)(r_1r_2)^2(\gamma_1{r_1}^4r_2(\gamma_1+\gamma_2)+\delta(\gamma_1{r_1}^2+\gamma_2{r_2}^2))}{\delta^3}
\end{align}
where the common term $\delta$ is given by
\begin{equation}
	\delta = \left((\gamma_1+\gamma_2)(r_1r_2)^2(\gamma_1{r_1}^2+\gamma_2{r_2}^2)\right)^{1/2}.
\end{equation}
\end{subequations}

\section{Experimental Setup}\label{sec:setup}
\begin{figure}[H]
	\centering
	\subfloat[]{
		\includegraphics[width=.5\linewidth]{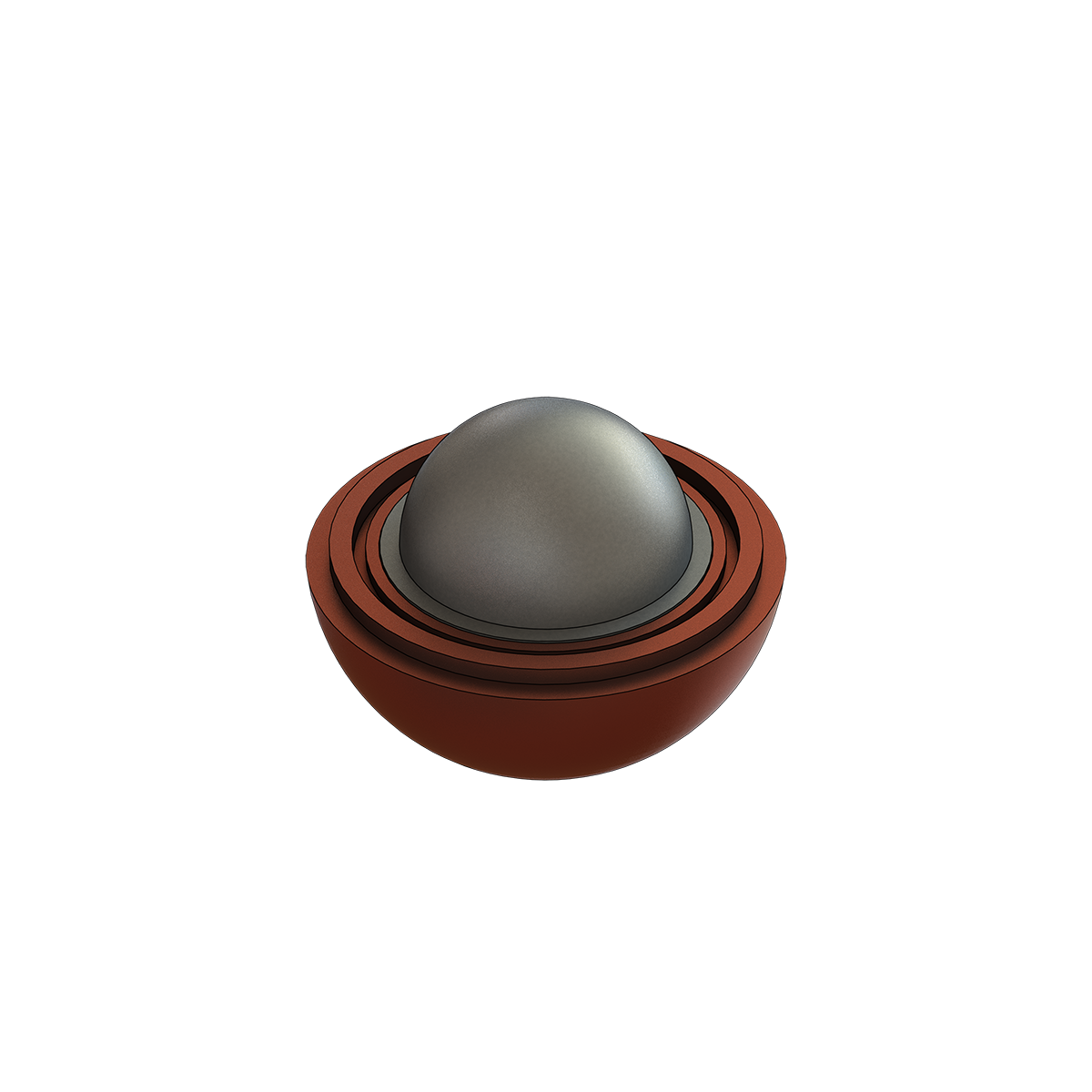}\label{fig:1}
	}
	\subfloat[]{
		\includegraphics[width=.5\linewidth]{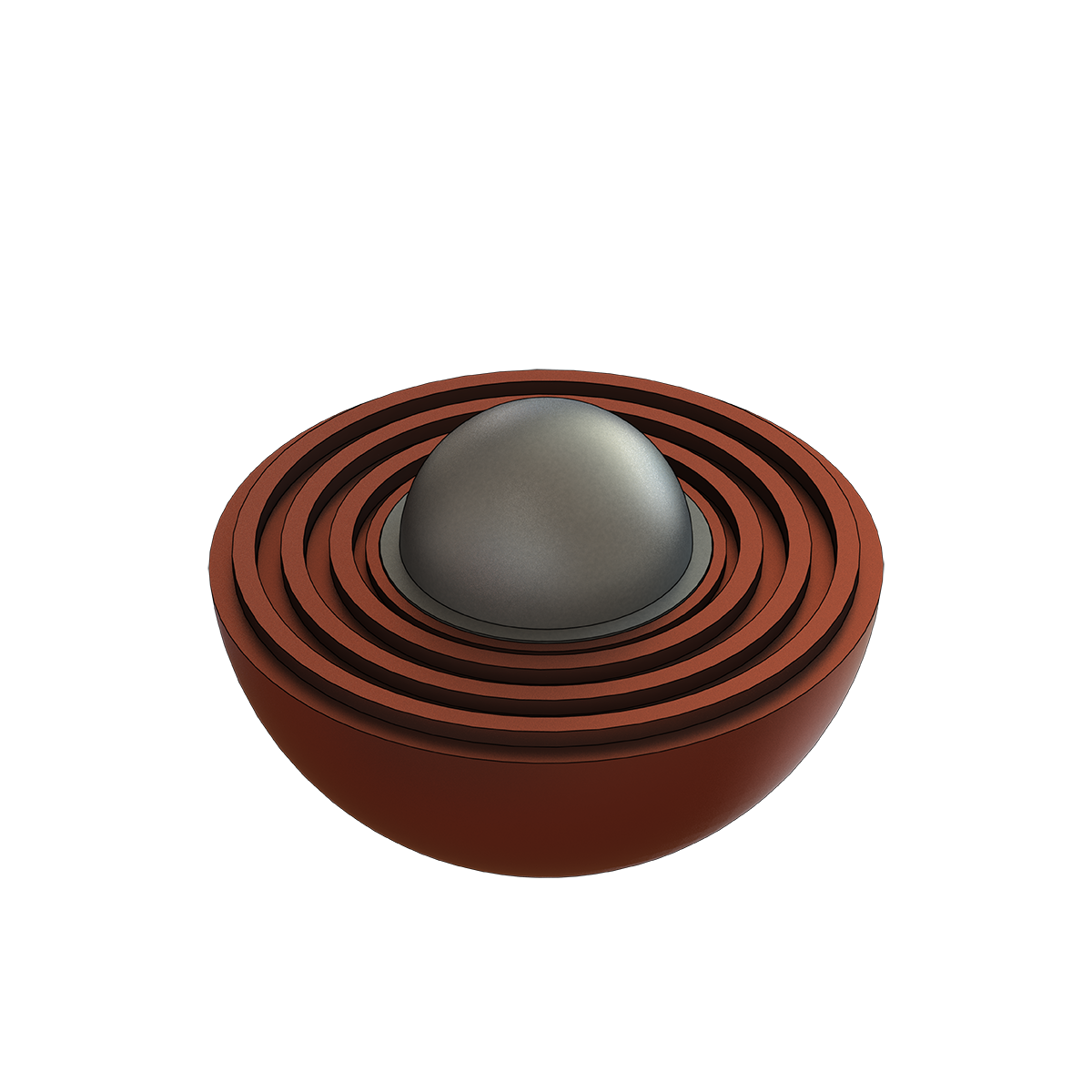}\label{fig:2}
	}
	\hspace{0mm}
	\subfloat[]{
		\includegraphics[width=.5\linewidth]{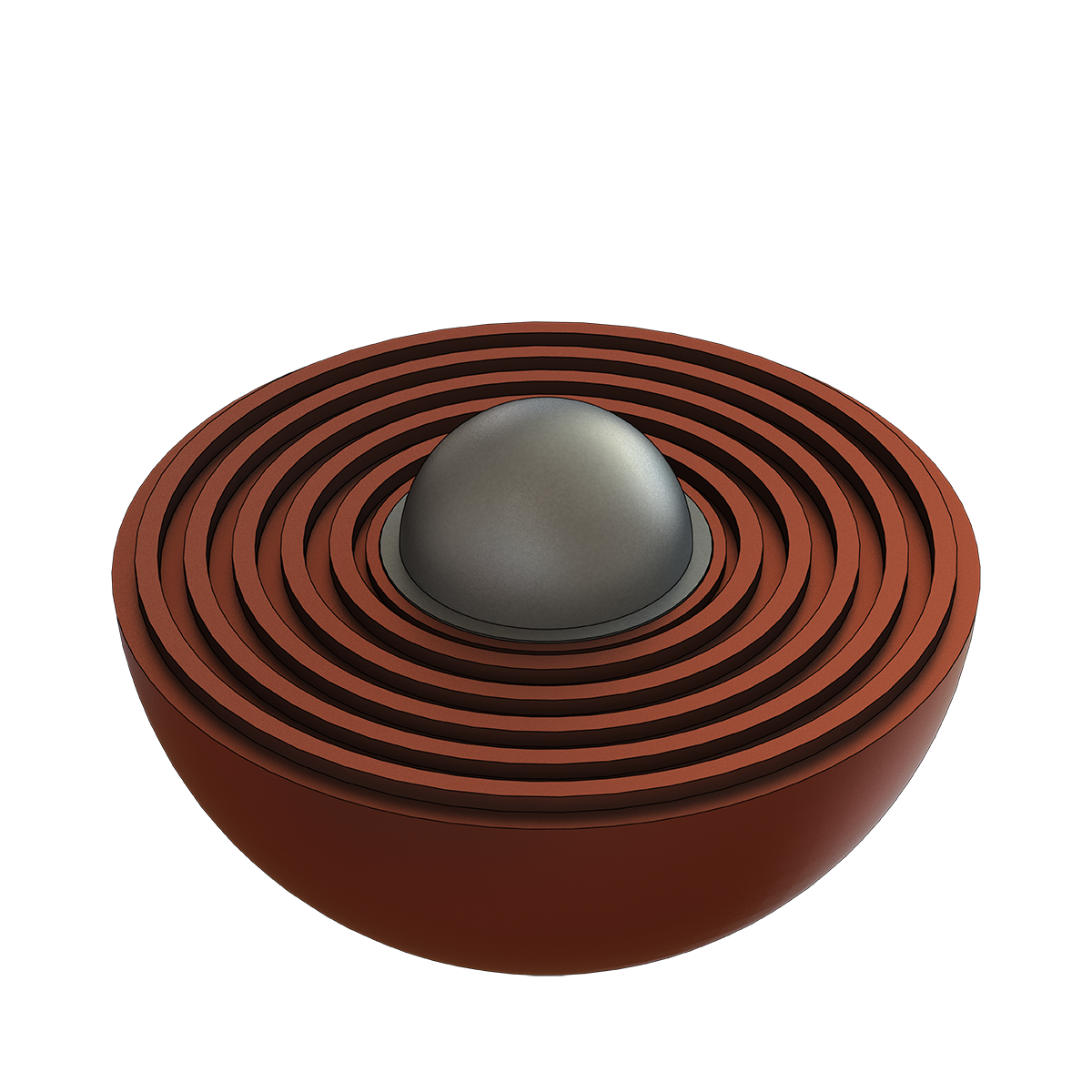}\label{fig:3}
	}
	\subfloat[]{
		\includegraphics[width=.5\linewidth]{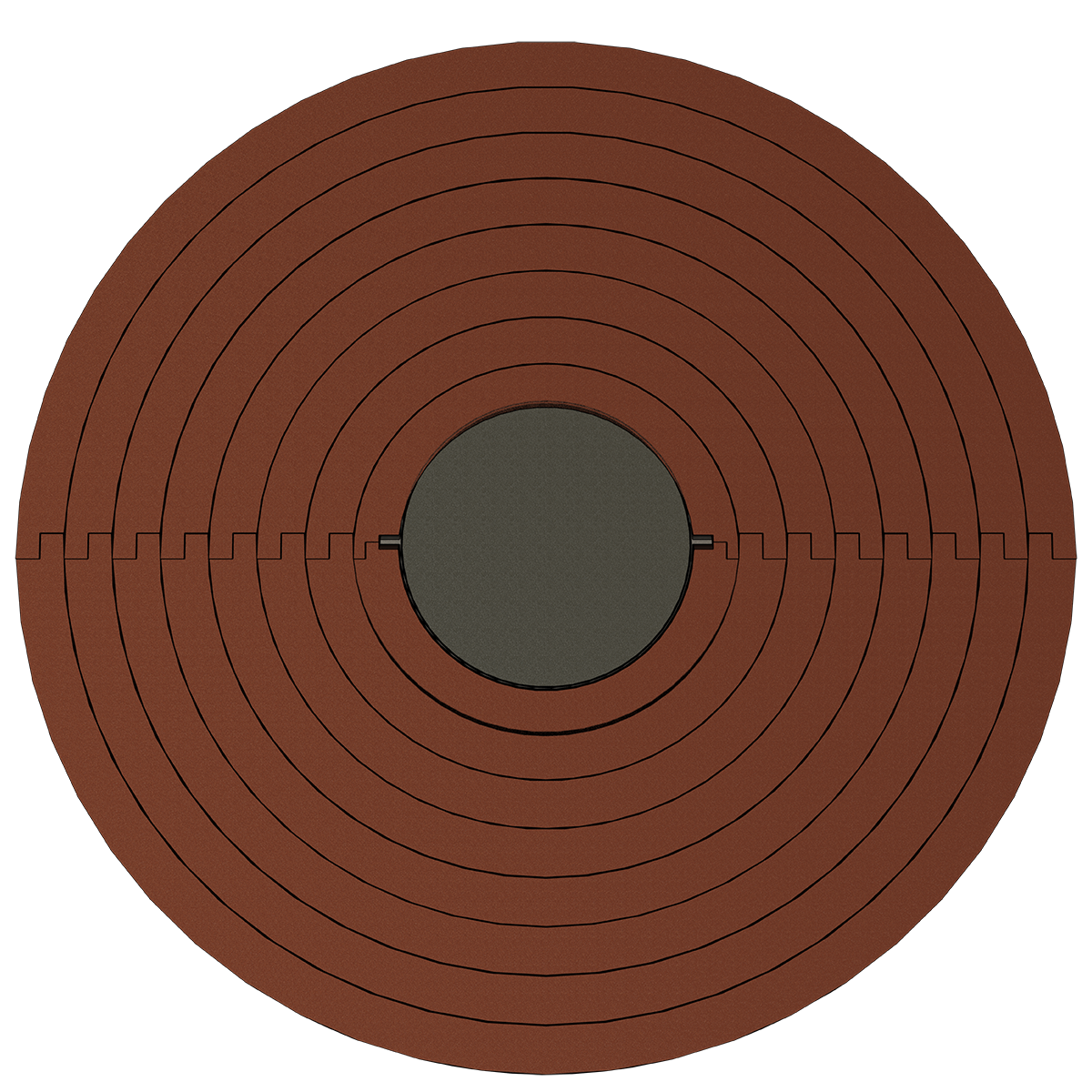}\label{fig:4}
	}
	\caption{(Subfigs.~\ref{fig:1},~\ref{fig:2}, and~\ref{fig:3}) three-dimensional renderings of the BeRP Ball reflected by 2.54, 5.08, and 7.62 cm of copper.  A two-dimensional engineering drawing of the 10.16-cm configuration detailing individual hemishells is shown in Subfig.~\ref{fig:4}.}
	\label{fig:configs}
\end{figure}

\begin{figure}[H]
	\centering
	\includegraphics[width=\linewidth]{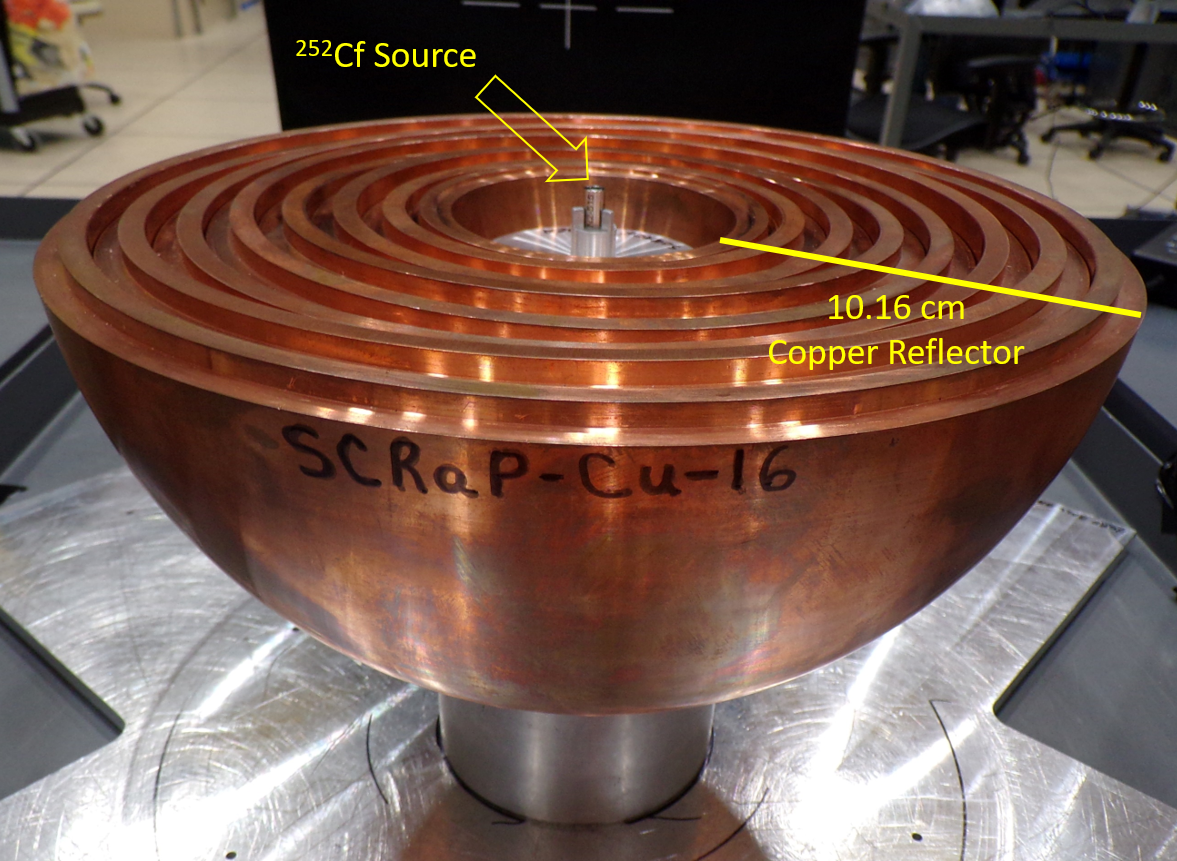}
	\caption{Photo of the $^{252}$Cf source in 10.16 cm of copper.}
	\label{fig:cf252_setup}
\end{figure}

Experimental data were obtained at the National Criticality Experiments Research Center within the Device Assembly Facility at the Nevada National Security Site to validate the two-region model. The fissile material was a 4.5-kg sphere of weapons-grade, alpha-phase plutonium encased in stainless steel (to prevent contamination) known as the BeRP Ball, which has been extensively detailed in integral benchmark experiments~\cite{BeRP_Mattingly2_2,BeRP_Ni1,BeRP_Ni2,BeRP_W,BeRP_benchmark_summary,scrap}. The BeRP Ball was reflected by various amounts of copper ranging from 1.27 cm to 10.16 cm in 1.27-cm increments for a total of eight configurations with a simulated $k_\text{eff}$ multiplication factor ranging between 0.8278 and 0.9394; these configurations are the same as the copper-only cases of the subcritical copper-reflected $\alpha$-phase plutonium (SCR$\alpha$P) benchmark.  Three-dimensional renderings with the bottom half of the hemishells and a two-dimensional schematic of the copper-reflected BeRP Ball assemblies are shown in Fig.~\ref{fig:configs}. The 10.16-cm copper configuration measurement was repeated with the BeRP Ball replaced by a $^{252}$Cf source; a photo of the open-face assembly is shown in Fig.~\ref{fig:cf252_setup}.  All measurements were 20 minutes long.

\begin{figure}[H]
	\centering
	\includegraphics[width=\linewidth]{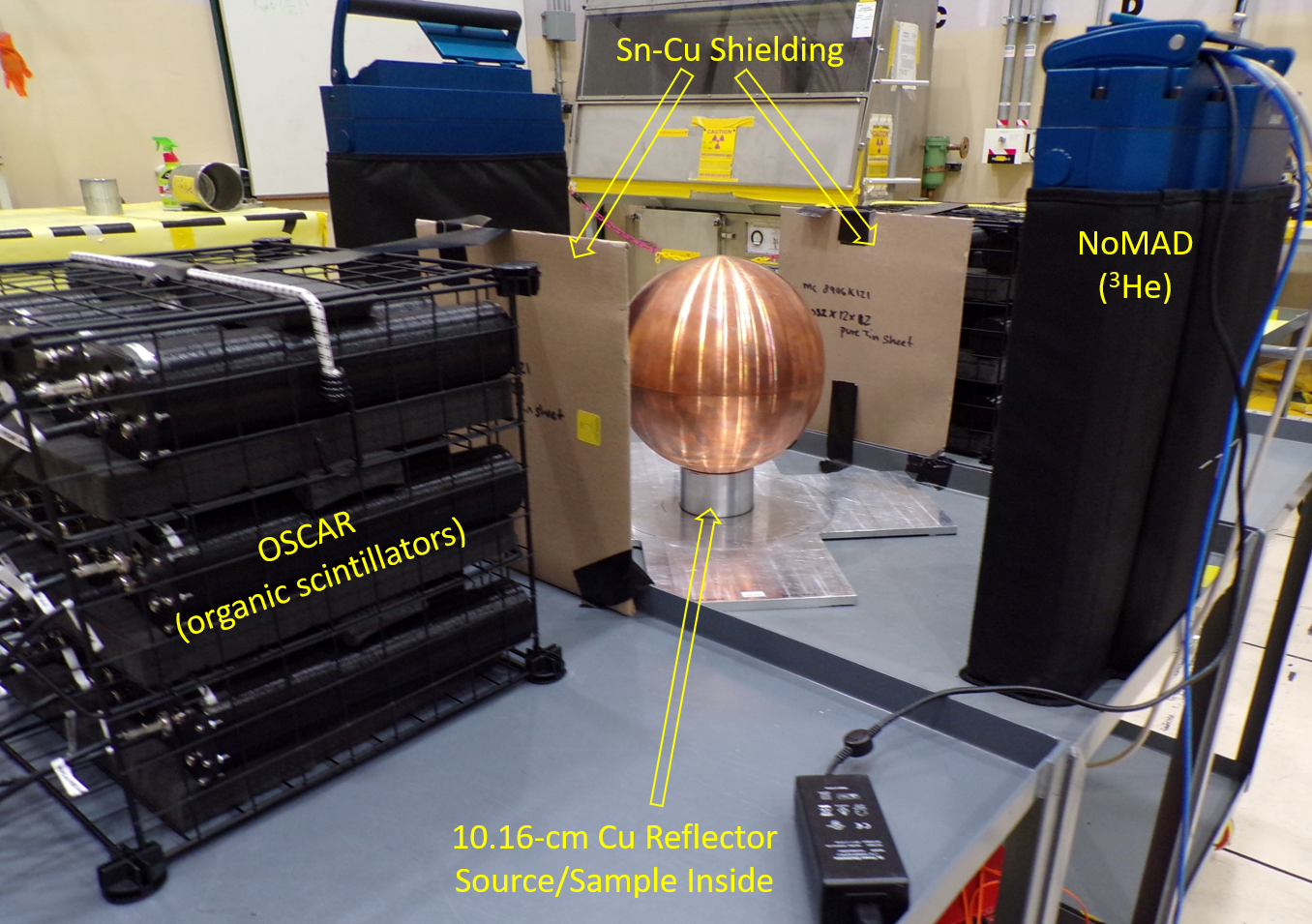}
	\caption{Annotated photo of the measurement setup including two organic scintillator arrays (OSCARs) and two Neutron Multiplicity $^3$He Array Detectors (NoMADs) all 47 cm from the center of the assembly.  The assembly comprises 10.16 cm of copper reflector.}
	\label{fig:setup}
\end{figure}

The assemblies were measured with an organic scintillator array (OSCAR~\cite{mikwa_validation}).  Previous works have also used organic scintillators to perform Feynman-alpha measurements~\cite{Feynman_Organics1,Feynman_Organics2,Feynman_Organics3}, although $^3$He detectors are traditionally used.   The OSCAR comprised 12, cylindrical, 5.08-cm thick $\times$ 5.08-cm diameter \textit{trans}-stilbene organic scintillators (C$_{14}$H$_{12}$) wrapped in polytetrafluoroethylene tape and housed in aluminum~\cite{stilbene,stilbene2}, coupled to photomultiplier tubes (PMTs), and arranged in a 3$\times$4 matrix. The face coupled to the PMT was optically polished and had a protective fused-silica window.  The detectors were contained in a wireframe mesh and held in place with porous polyurethane foam (0.021 g/cm$^3$).  The front face of the array was 47 cm from the center of the assembly.  The detection threshold was 35 keVee and tin-copper graded shielding was placed between the assembly and the OSCAR to preferentially shield 60-keV gamma rays from the build-up of $^{241}$Am in plutonium samples. The detectors were read out by a digitizer (CAEN, v1730) with a 500-MHz sampling rate (2-ns sampling period) in 288 ns acquisition windows per pulse.  The detection system is identical to that of Refs.~\cite{RA_w_Organics,mikwa_validation}, which contain further details.  A second OSCAR and two Neutron Multiplicity $^3$He Array Detectors (NoMADs -- 15 $^3$He tubes embedded in a high-density polyethylene matrix~\cite{scrap}) were also present, all 47 cm from the center of the assembly, but data obtained from these systems are not used in this work.  A photo of the experimental setup is shown in Fig.~\ref{fig:setup}.

\section{Data Analysis}\label{sec:analysis}
The data analysis is performed in two steps: the analysis of raw data to obtain the list mode data (a list of neutron detection times) and the Feynman-alpha analysis that creates and fits a Feynman histogram from the list mode data. The former is discussed in subsec.~\ref{sec:analysis1} and the latter in subsec.~\ref{sec:analysis2}.

\subsection{Detector Analysis to Obtain List Mode Data}\label{sec:analysis1}
\begin{figure}[H]
	\centering
	\includegraphics[width=\linewidth]{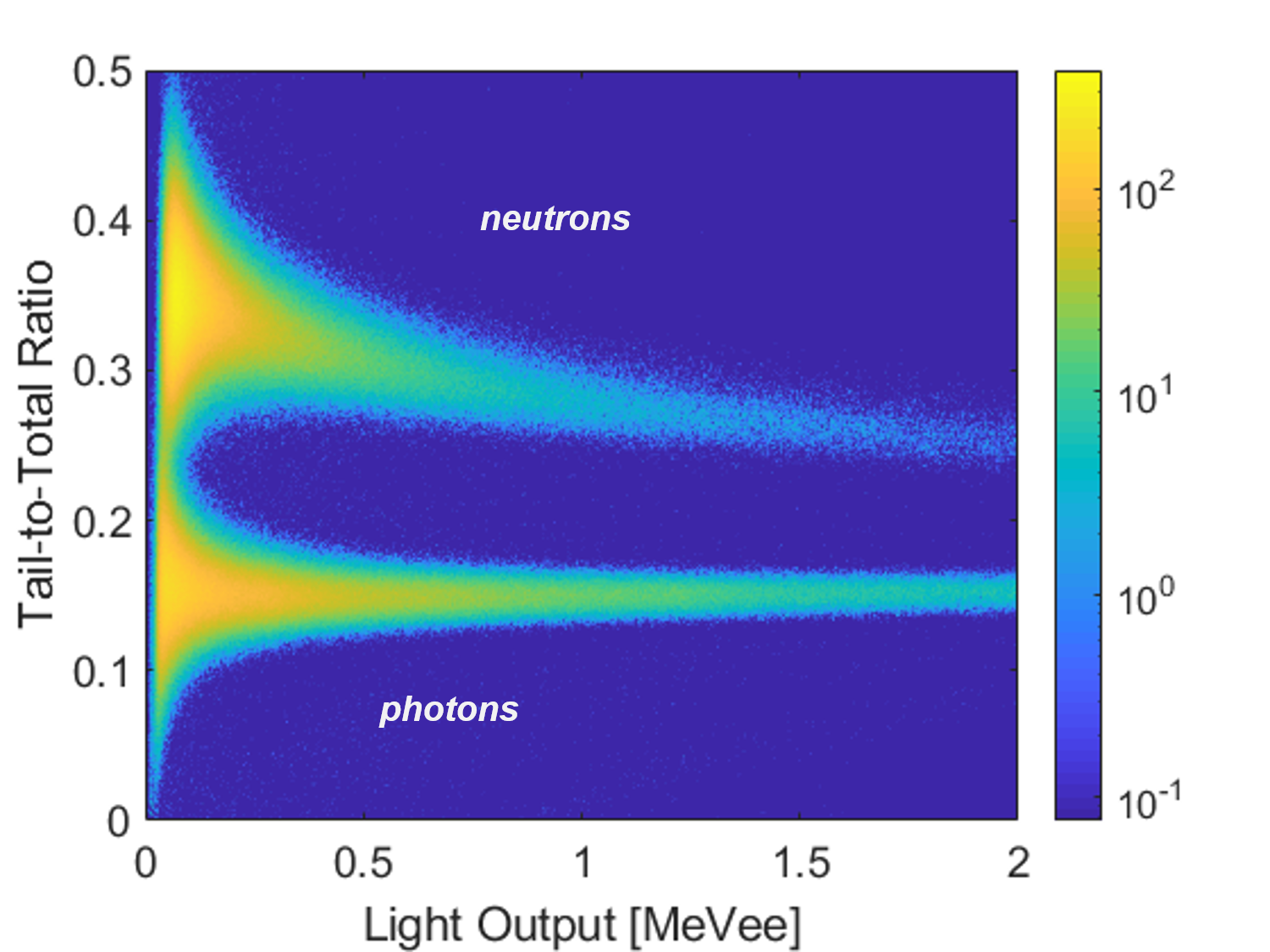}
	\caption{Pulse shape discrimination plot for the copper-reflected $^{252}$Cf measurement.}
	\label{fig:psd}
\end{figure}

Preprocessing is performed to obtain clean pulses; clipped pulses (those that exceed 2.39 MeVee) and pile-up pulses (multiple pulses in the same 288-ns pulse window) are discarded.  Since organic scintillators are sensitive to both neutrons and photons, a charge-integration-based pulse shape discrimination (PSD) analysis is performed to obtain the neutrons only~\cite{PSD1,PSD2}.  The PSD algorithm is performed for each minute of measurement, for each detector, and as a function of energy~\cite{PSD_Slice}; a sample PSD plot for the $^{252}$Cf measurement is shown in Fig.~\ref{fig:psd}. The times-of-detection associated with clean neutron pulses are sorted and therein the list mode data is obtained.

\subsection{Feynman-$\alpha$ Analysis}\label{sec:analysis2}
\begin{figure}[H]
	\centering
	\includegraphics[width=\linewidth]{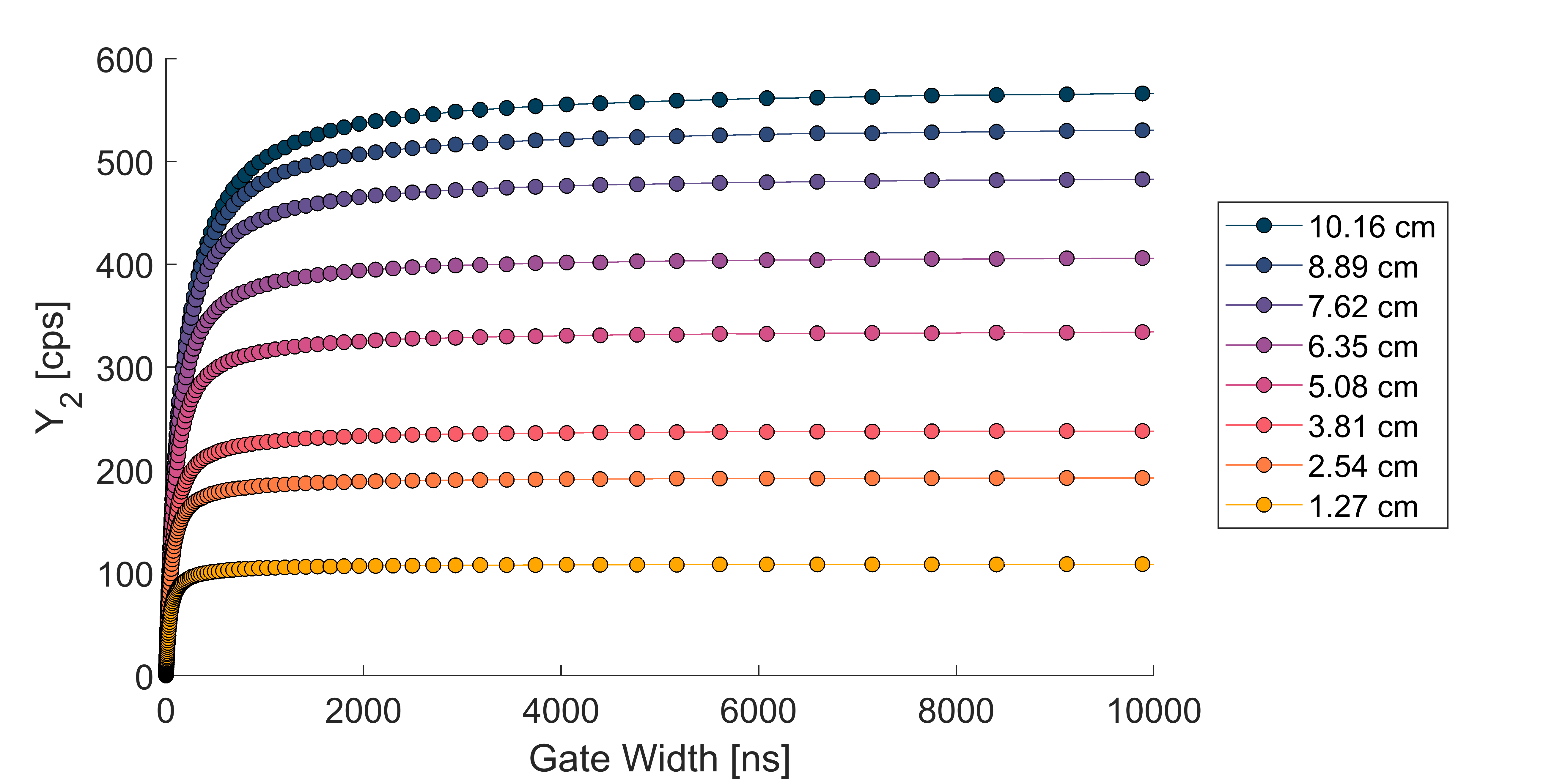}
	\caption{Feynman histograms for the BeRP Ball reflected by various amounts of copper.  Error bars (one standard deviation) are smaller than the markers.}
	\label{fig:Feynman_BeRP}
\end{figure}

\begin{figure}[H]
	\centering
	\includegraphics[width=\linewidth]{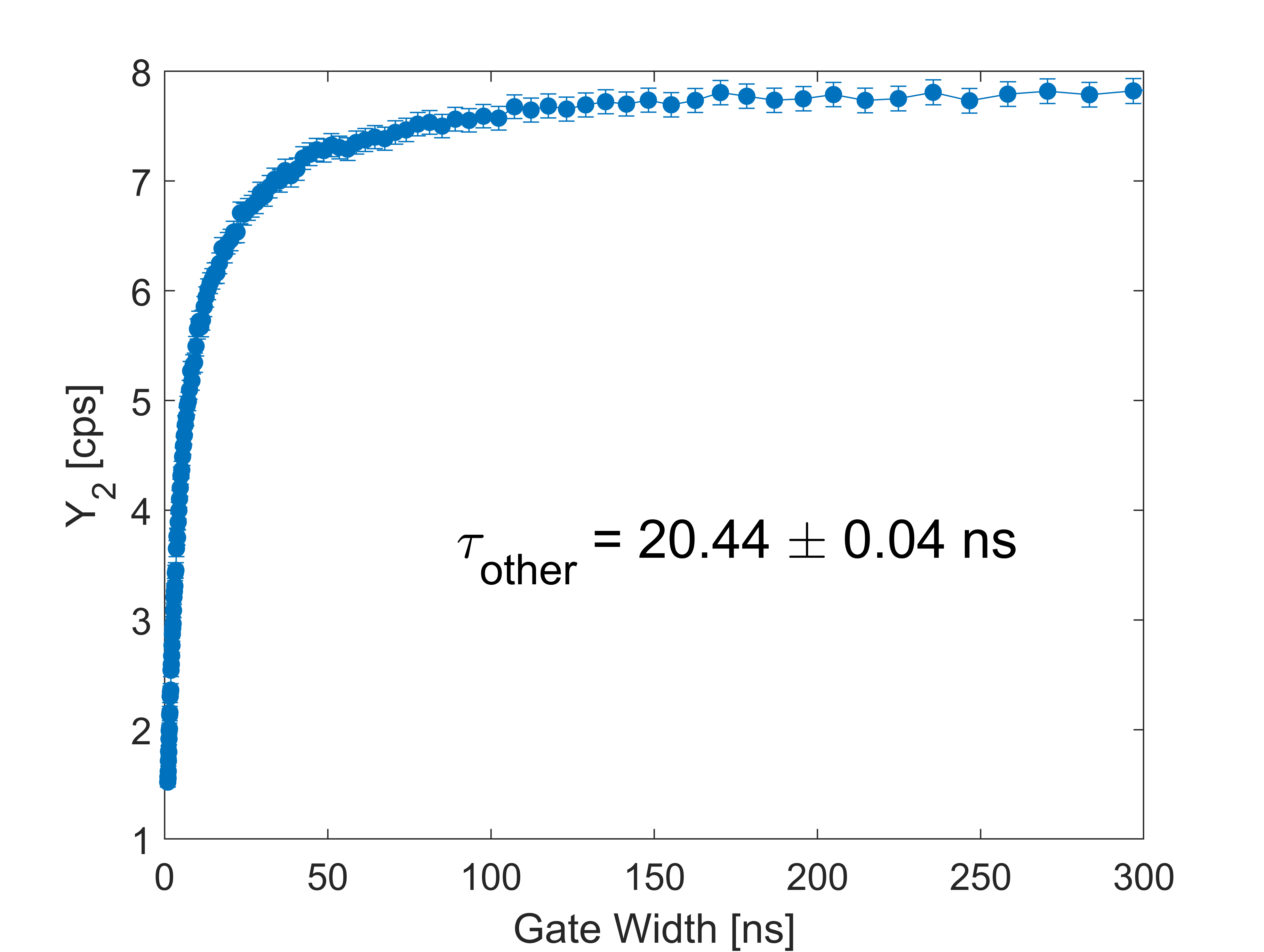}
	\caption{Feynman histogram and fit value for the measurement of $^{252}$Cf in 10.16 cm of copper.}
	\label{fig:Feynman_Cf}
\end{figure}

The Feynman histograms are constructed by calculating $Y_2$ from Eqn.~\eqref{eq:Y_moment} as a function of $\tau$.  The analysis of the BeRP Ball data used 200 values of $\tau$ logarithmically distributed between $10^2$ and $10^4$ ns, and the analysis of the $^{252}$Cf data used 200 values of $\tau$ distributed between $10^1$ and $10^4$.  The resultant histograms are shown in Figs.~\ref{fig:Feynman_BeRP} and~\ref{fig:Feynman_Cf}, respectively.  The histograms are fit by Eqn.~\eqref{eq:Y2} with a nonlinear least squares algorithm and weighting, with weights determined by the analytic method.  The histograms were also fit with the one-exponential model in Eqn.~\eqref{eq:Ya_one} for comparison.

\section{Results and Discussion}\label{sec:results}
The two-region Feynman-alpha model is validated by comparing measured values of the prompt neutron period to simulated reference values for identical measurements given in Ref.~\cite{mikwa_validation}; Ref.~\cite{mikwa_validation} validated the two-region Rossi-alpha model.  The simulated values from MCNP$\textregistered$\footnote{MCNP$\textregistered$ and Monte Carlo N-Particle$\textregistered$ are registered trademarks owned by Triad National Security, LLC, manager and operator of Los Alamos National Laboratory. Any third party use of such registered marks should be properly attributed to Triad National Security, LLC, including the use of the designation as appropriate. For the purposes of visual clarity, the registered trademark symbol is assumed for all references to MCNP within the remainder of this paper.} represent an independent determination of the prompt neutron periods, not a simulation of the detector response.  Three parameters were obtained from simulation: $k_\text{eff}$, $\beta_\text{eff}$ (the effective delayed neutron fraction), and $\tau_c$ (the mean lifetime of a neutron in the fissile core region). The KCODE subroutine of MCNP was used to calculate $k_\text{eff}$ and $\beta_\text{eff}$ while two F4 track-length tallies weighted by inverse velocity were used to calculate $\tau_c$ in the SDEF subroutine. The prompt neutron period was then obtained from $k_\text{eff}$, $\beta_\text{eff}$, and $\tau_c$.

Raw Feynman-alpha analysis produced prompt periods that were greater than the reference values by a constant time offset; the uniform mean difference was 21.37 ns and the inverse-variance-weighted mean difference was 20.13 ns. The time offset is due to lifetimes not associated with the multiplication kinetics such as detector dead time or neutron cross talk wherein one neutron registers multiple detections by scattering in multiple detectors~\cite{cross_talk,Feynman_crosstalk}.  

The mean lifetime due to non-multiplication kinetics, $\tau_\text{other}$, is determined by measuring a $^{252}$Cf source and repeating Feynman analysis; samples are typically small powders in which multiplication is negligible.  The $^{252}$Cf measurement in this work for the 10.16-cm copper reflector case yielded $\tau_\text{other} = 20.44\pm0.04$ ns, as annotated in Fig.~\ref{fig:Feynman_Cf}. Similar measured data for the other configurations are not available, so the measured $\tau_\text{other}$ is verified with simulated results. The measurement was simulated using MCNPX-PoliMi and included all detection systems, tables, stands, shielding, and the floor.  The simulation models detector response for 20-minute measurements of the 2.54, 5.08, 7.62, and 10.16 cm cases, and respective values are shown in Tab.~\ref{tab:tau}. The mean non-multiplicative time constant from simulation is $\langle \tau_\text{other}\rangle_\text{sim} = 19.83 \pm 0.45$ ns, consistent with the measured value. A measured value for each configuration is generally preferred; in this work, the measured $\tau_\text{other}$ will be used for all configurations since there is no apparent trend shown in the simulated values.

\begin{table}[H]
	\footnotesize
	\centering
	\caption{Simulated $\tau_\text{other}$ values for the 2.54, 5.08, 7.62, and 10.16 cm cases and a measured value for the 10.16 cm case.}
	\begin{tabular}{rcc}
		\toprule
		\multicolumn{1}{l}{Cu Thickness [cm]} & Simulated $\tau_\text{other}$ [ns] & Measured $\tau_\text{other}$ [ns] \\
		\midrule
		2.54  & 20.00 $\pm$ 0.03 & -- \\
		5.08  & 18.44 $\pm$ 1.79 & -- \\
		7.62  & 19.88 $\pm$ 0.11 & -- \\
		10.16 & 21.02 $\pm$ 0.03 & 20.44 $\pm$ 0.04 \\
		\bottomrule
		\bottomrule
	\end{tabular}%
	\label{tab:tau}%
\end{table}%

\begin{figure}[H]
	\centering
	\includegraphics[width=\linewidth]{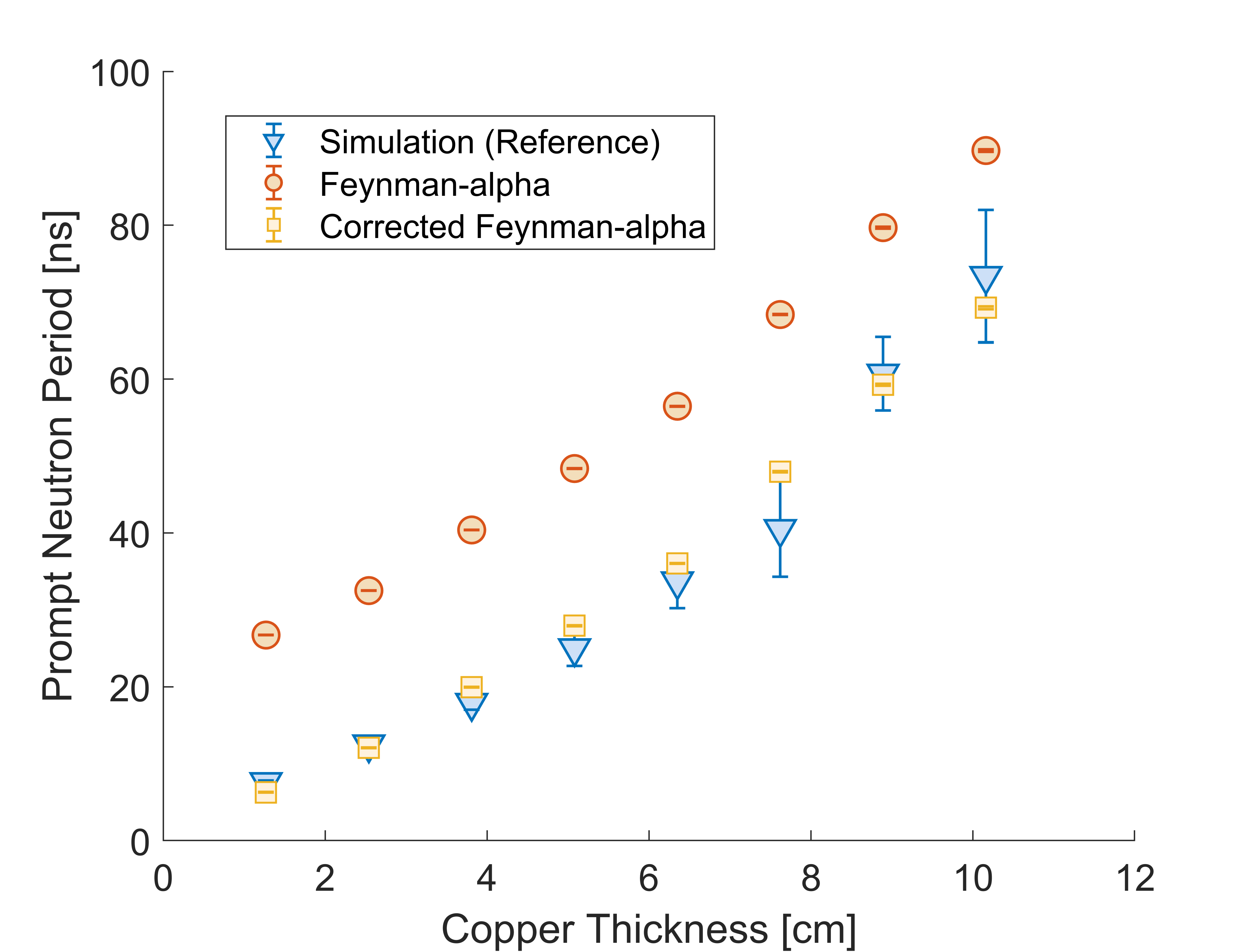}
	\caption{Comparison of measured and simulated prompt neutron periods, treating the simulated values as reference. The `Feynman-alpha' data are pre-correction, whereas the `Corrected Feynman-alpha' data are subtracted by the non-multiplication time constant determined from analysis on non-multiplying $^{252}$Cf data. }
	\label{fig:pre_val}
\end{figure}

The raw Feynman-alpha, corrected Feynman-alpha (raw values minus $\tau_\text{other}$), and simulated reference values are shown in Fig.~\ref{fig:pre_val}. The values are also summarized and compared to the Rossi-alpha values in Tab.~\ref{tab:addlabel}. There is excellent absolute agreement between the corrected Feynman-alpha and reference values; additionally, the Feynman-alpha values are more precise than the Rossi-alpha values (by one to two orders of magnitude) with one-standard-deviation measurement uncertainties less than 1\%. The Feynman-alpha values are more accurate than the Rossi-alpha values below $k_\text{eff}\approx0.92$, while the opposite is generally true above; the error for the more-accurate prompt neutron period estimate for each assembly is presented in blue in Tab.~\ref{tab:addlabel}.

\begin{figure}[H]
	\centering
	\includegraphics[width=\linewidth]{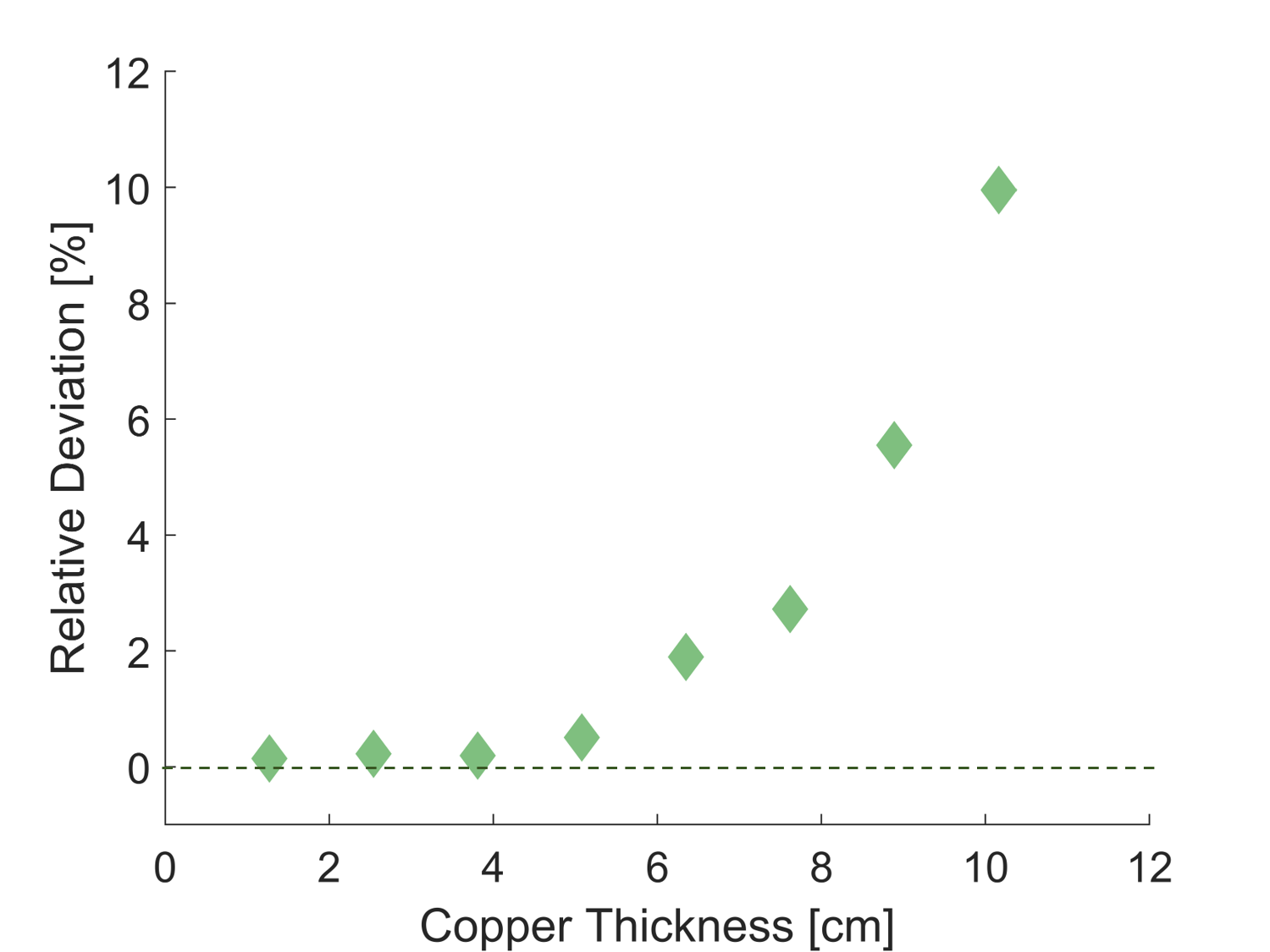}
	\caption{Relative deviation of the one-region model estimate of the prompt neutron period from that of the two-region model.}
	\label{fig:rel_dev}
\end{figure}

The now-validated two-region model is compared to the one-region model to determine regimes of applicability.  Note that the two-region model is a generalization of the one-region model and that $R$ in Eqn.~\eqref{eq:alpha} reduces to zero or unity when reflector is negligible~\cite{mikwa_validation}. Therefore, the two-region model can always be used and is not prohibitively more computationally expensive. The relative deviation between the one- and two-region models ($(\alpha_1-\alpha_2)/\alpha_2$) is shown in Fig.~\ref{fig:rel_dev}.  It is shown that the models deviate by more than 10\% as the reflector and reflection increase, and are nearly identical when there is only a small amount of reflector.

\begin{table*}[]
	\footnotesize
	\centering
	\caption{Prompt neutron period values for the Feynman-alpha, Rossi-alpha, and simulation approaches with  one-standard-deviation uncertainties. The $\text{error}=(\text{measured}-\text{simulated})/(\text{simulated})$ for the measured approaches is given and the error for the more accurate measured approach is {\color{blue}displayed in blue} for each configuration.  The simulated and Rossi-alpha values come from Ref.~\cite{mikwa_validation}.}
	\begin{tabular}{r|rr|rr|rr}
		\toprule
		\multicolumn{1}{c|}{Assembly} & \multicolumn{2}{c|}{Simulation} & \multicolumn{2}{c|}{Feynman-alpha} & \multicolumn{2}{c}{Rossi-alpha} \\
		\multicolumn{1}{c|}{Cu Thickness [cm]} & \multicolumn{1}{c}{$k_\text{eff}$} & Prompt Period [ns] & Prompt Period [ns] & \multicolumn{1}{c|}{Error [\%]} & Prompt Period [ns] & \multicolumn{1}{l}{Error [\%]} \\
		\midrule
		1.27  & 0.8278 & 7.6 $\pm$ 0.2 & 6.28 $\pm$ 0.04 & {\color{blue}-17\%} & 13.4 $\pm$ 1.0 & 76\% \\
		2.54  & 0.8604 & 12.5 $\pm$ 0.4 & 12.06 $\pm$ 0.05 & {\color{blue}-3\%}  & 19.5 $\pm$ 1.0 & 57\% \\
		3.81  & 0.8831 & 17.9 $\pm$ 0.9 & 19.95 $\pm$ 0.06 & {\color{blue}11\%}  & 27.6 $\pm$ 4.0 & 54\% \\
		5.08  & 0.9005 & 25.0 $\pm$ 2.3 & 27.93 $\pm$ 0.05 & {\color{blue}12\%}  & 32.1 $\pm$ 4.9 & 28\% \\
		6.35  & 0.9137 & 33.7 $\pm$ 3.5 & 36.03 $\pm$ 0.05 & {\color{blue}7\%}   & 40.8 $\pm$ 7.4 & 21\% \\
		7.62  & 0.9239 & 40.5 $\pm$ 6.2 & 47.95 $\pm$ 0.07 & 18\%  & 43.6 $\pm$ 5.1 & {\color{blue}8\%} \\
		8.89  & 0.9322 & 60.7 $\pm$ 4.8 & 59.25 $\pm$ 0.13 & {\color{blue}-2\%}  & 68.8 $\pm$ 3.6 & 13\% \\
		10.16 & 0.9394 & 73.4 $\pm$ 8.6 & 69.27 $\pm$ 0.17 & -6\%  & 75.6 $\pm$ 4.5 & {\color{blue}3\%} \\
		\bottomrule
		\bottomrule
	\end{tabular}%
	\label{tab:addlabel}%
\end{table*}%

\section{Summary and Conclusion}\label{sec:conclusion}
The two-region Feynman-alpha model was derived from the two-region Rossi-alpha model, rigorous propagation of measurement uncertainty was developed, and the two-region Feynman-alpha model was validated with organic scintillator measurements of copper-reflected, weapons-grade plutonium.  The uncertainty propagation in this work should be used to improve fit accuracy and to properly propagate uncertainty, as demonstrated by other works~\cite{mikwa_unc,mikwa_ANS_unc,mikwa_ANS}. Having validated the two-region model and demonstrated accuracy over the one-region model, the two-region model should be used over the one-region model.  If the two-region model is unnecessary, it will reduce to the traditional case of the one-region model. In special circumstances, the two-region model may find two $\alpha$ eigenmodes if they dominate reflector-induced modes; modal effects are the subject of future work.

The copper-reflected plutonium measurements were complemented by a $^{252}$Cf measurement to correct for non-multiplicative time correlations such as neutron cross talk or dead time. The method was effective since the $^{252}$Cf is non-multiplying, thus this approach is recommended. One $^{252}$Cf measurement was used as a representative for all cases and copper thicknesses since simulated values for the other cases were similar. Future work will further study the corrective $^{252}$Cf measurement and investigate alternative cross talk and non-multiplicative corrections.

The Feynman-alpha approach offers better precision than the Rossi-alpha approach as well as improved accuracy for $k_\text{eff}<0.92$, whereas the Rossi-alpha method is generally more accurate for larger multiplications.  The trends are expected since the Feynman-alpha approach is an integral of the Rossi-alpha approach, though the Feynman-alpha approach is expected to struggle as $k_\text{eff}$ approaches unity (whereas the Rossi-alpha approach improves as $k_\text{eff}$ approaches unity).  More fission chains overlap as $k_\text{eff}$ increases, which can obfuscate accidentals and associated uncertainties in the Feynman-alpha method.  Therefore, there is an optimization between Feynman-alpha and Rossi-alpha for regimes of preference, which is the subject of future work through the upcoming Measurement of Uranium Subcritical and Critical (MUSiC) benchmark~\cite{MUSiC_mcspaden,MUSiC_weldon_ANS_2020}.

\section*{Acknowledgments}
This work was partially supported by the  National Science Foundation Graduate Research Fellowship under Grant No. DGE-1256260, the Consortium for Verification Technology under Department of Energy National Nuclear Security Administration award number DE-NA0002534, the Consortium for Monitoring, Technology, and Verification under Department of Energy National Nuclear Security Administration award number DE-NA0003920, and the DOE Nuclear Criticality Safety Program, funded and managed by the National Nuclear Security Administration for the Department of Energy.  Any opinion, findings, and conclusion or recommendations expressed in this material are those of the authors and do not necessarily reflect the views of any funding organization.  

\bibliographystyle{elsarticle-num} 
\bibliography{bibliography}
\end{document}